# Evaluating the Usefulness of Sentiment Information for Focused Crawlers

**Tianjun Fu[1], Ahmed Abbasi[2], Daniel Zeng[1], Hsinchun Chen[1]**
University of Arizona[1], University of Wisconsin-Milwaukee[2]
futj@email.arizona.edu, abbasi@uwm.edu, zeng@email.arizona.edu, hchen@eller.arizona.edu

**Abstract:**
Despite the prevalence of sentiment-related content on the Web, there has been limited work on focused crawlers capable of effectively collecting such content. In this study, we evaluated the efficacy of using sentiment-related information for enhanced focused crawling of opinion-rich web content regarding a particular topic. We also assessed the impact of using sentiment-labelled web graphs to further improve collection accuracy. Experimental results on a large testbed encompassing over half a million web pages revealed that focused crawlers utilizing sentiment information as well as sentiment-labelled web graphs are capable of gathering more holistic collections of opinion-related content regarding a particular topic. The results have important implications for business and marketing intelligence gathering efforts in the Web 2.0 era.

## 1. Introduction

Content expressing user opinions such as online reviews, recommendations, blog articles, and discussion forums has been proliferating in a wide array of Web 2.0 and social media applications. Such content often contains obvious *sentiment polarity* (negative/positive/neutral). Modern business and marketing intelligence gathering efforts seek to collect and analyze Web content that is not only topically relevant and also provides insights regarding the opinions of various stakeholder and special interest groups [10]. For instance, companies are increasingly interested in how they are perceived by environmental and animal rights groups. Many individuals, including researchers, also seek content published by people that share the same sentiment polarity on a topic, resulting in the phenomenon known as Cyber Balkanization [1]. However, limited effort has been made to develop Web crawling methods for sentiment related data. Previous crawler studies primarily emphasized the collection of topic-relevant content; these approaches perform poorly when applied to sentiment-rich text [2].

In this study we propose a focused crawler algorithm that collects topic and sentiment relevant Web content. Sentiment polarity information of each collected Web page, together with its topic information, are used to navigate the crawler. Moreover, the sentiment information, coupled with web graph-based information, helps the crawler learn and apply appropriate strategies when relevant content is near *irrelevant* or *off-topic* Web pages.

## 2. Previous Work and Research Questions

Focused crawlers use the available context information to guide the navigation of links with the goal of efficiently locating highly relevant target pages [3] and are seen as a way to address the scalability limitations of universal search engines. According to Pants & Srinivasan [4], there are three types of contextual information useful for estimating the benefit of following a link: link context, ancestor pages, and web graph. Link context refers to the lexical content around the link, which can range from anchor text to the whole content of the link's parent page. Ancestor pages are the lexical content of the link's ancestor pages. Web graph refers to the hyperlink graph comprised of in and out links between web pages.

Most focused crawler methods utilize link context and ancestor pages, and focus on identifying the embedded topic information. To collect Web content on a topic with a specific sentiment polarity, it is still possible to use those methods: One can first collect all the topic relevant content with any sentiment polarity and then select targeted content by identifying the sentiment polarities. However, this approach greatly diminishes the main advantages of focused crawlers: efficient use of computing time and storage space..

Sentiment analysis, which aims to identify the polarity and intensity of text with respect to certain topic, has become a popular research field in recent years [11, 12]. One way is to evaluate the polarity based on a set of selected keywords which have strong polarity biases. To the best of our knowledge, sentiment information has never been utilized by previous crawler studies. In our study, we aim to answer the following research question:

*Q1: Is sentiment information useful in crawling web content of a topic with specific sentiment polarity?*

Prior research has noted that typical focused crawlers lack appropriate learning strategies when relevant pages are near off-topic pages [5]. Most crawlers ignore links on irrelevant pages and only follow links on relevant pages. This approach is very likely to cause the crawlers to run into local maximum points and miss relevant content that is often just a few steps away. This problem is exacerbated in the era of Web 2.0 due to the sheer volume of online communication  as well as the ever-growing number of communication modes Appropriate learning strategies must be capable of effectively evaluating irrelevant pages' potential for leading to relevant pages.

Among the three context information mentioned [4], web graph relies least on the lexical content of a page, making it most suitable to solve the above problem. However, very few crawler studies have explored web graph due to the limited available graph information. Chakrabarti et al. [6] tried to identify hub nodes in web graphs while Chau & Chen [7] modelled the web graph as a weighted, single-layer neutral. These studies used a web graph to pass accumulated weights to children pages. But labelled graphs, where nodes are enriched with additional information (e.g., topic and sentiment relevance), have rarely been considered in prior crawler research. For typical focused crawlers, nodes in a web graph only have two classes: topically relevant and irrelevant. The number of node classes is too small to be used. However, if sentiment information of a node is also considered, the number of node classes will be increased to four (sentiment relevant/irrelevant) or even six (positive, negative, neutral), making class information possible to be explored. This leads to our second research question:

*Q2: Can labelled web graphs, leveraging node sentiment information, improve crawlers' ability to learn strategies for collecting relevant pages that are near irrelevant ones?*

## 3. System Design

To answer these research questions raised in the previous section, we have developed a new focused crawler that can leverage sentiment information. This graph-based sentiment crawler consists of four modules: crawler, queue management, text classifier, and graph comparison. The first two modules are common for focused crawlers. The crawler module crawls any URL popped by the queue management module. The queue management module ranks the entire candidate URLs according to their weights. Weights of candidate URLs are determined by the last two modules, described in detail below.

### 3.1 Text Classifier Module

Our text classifier module consists of a topic classifier and a sentiment classifier. A set of training data that contain relevant web pages examples (both in topic and sentiment) and irrelevant web pages are used to train the two classifiers. Each classifier adopts a similar simple, computationally efficient categorization approach suitable for use within a focused crawler. Topic and sentiment related word n-gram features are learned from the training corpus using the information gain heuristic, where each selected feature is weighted based on its level of entropy reduction [11, 12]. Each test page's topic relevance score is computed as the weighed sum of the present features. The page is considered relevant if its score is above a pre-defined topic threshold. A similar approach is used for sentiment classification, using the sentiment features and a sentiment threshold. More advanced sentiment analysis methods will be explored in the future.

By applying the text classifier module, each collected web page is categorized in the following four classes:

| | |
|---|---|
| C1: Relevant topics and sentiment | C2: Relevant topics only |
| C3: Relevant sentiment only | C4: Irrelevant topics and sentiment |

Only C1 pages are considered targets of our system. Previous studies have also shown the benefit of exploring outlinks of targeted web pages so that it is nature for us to assign highest weights for outlinks of C1 pages. For outlinks of other types of pages (C2, C3, C4), their weights are either calculated by the graph comparison module described in section 3.2 or set to predefined values discussed in section 4.2.

### 3.2 Graph Comparison Module

The graph comparison module intends to explore the usage of node class information of a web graph in learning strategies for irrelevant pages, described in Q2. The following example illustrates why we need to develop this module. Suppose the following path leads to a targeted C1 page: C1→C2→C4→C1 (target). The labels represent the classes of pages on the path. A typical focused crawler will explore all the outlinks of the first C1 page and therefore collect page C2. If it is a traditional topic-based focused crawler, it will further advance along the path sometime and collect page C4. Since the C4 page is neither topic relevant nor sentiment relevant, the crawler will not be interested to explore this path anymore and miss our targeted C1 page. To evaluate the value of this C4 page, we cannot rely on its lexical content. However by looking at the path that leads to C4 (C1→C2→C4), we notice that this path is in fact a part of the path that leads to our target C1 page. Similarly, the web graph of C4 consisting of multiple paths is in fact a sub-graph of that of the targeted C1 pages. Therefore the similarity of web graphs can be used to estimate how close an irrelevant page is to relevant content.

Algorithmically, our graph comparison module calculates weights of irrelevant pages by comparing their identified web graphs with those of training data. Each web graph is represented by its *random walk paths* (RWP). A RWP is a path consisting of sequence of node classes with a probability. At each step, a random walk either jumps to one of the neighbours or stops based on a probability distribution. For example, in a 3-level full binary tree with 0.1 stop probability and equal "jump" probability, a 3-node RWP originated from the root node (level 0) has the probability of 0.45*0.45*0.1.

Similarity between RWPs is measured by their *Levenshtein distance*. The similarity between two web graphs is the aggregated value of similarities among their RWPs multiplied by RWPs' possibilities. We define the weight of a candidate URL as the *ratio* of its web graph's average similarity score with web graphs of the relevant training data to the average similarity score with those of the irrelevant training data.

Two issues need to be addressed for our graph comparison module. First, the efficiency of the graph comparison module needs to be considered. Limitations on the graph levels and number of nodes in a web graph should be applied. Calculations on web graphs of training data should be done during the training process. Second, the web graph of a candidate URL can be updated during the crawling process when new ancestor pages or inlink pages are discovered. Therefore the weights of candidate URLs should also be updated from time to time. Currently we update the weights every time we have collected a predefined number of irrelevant pages.

## 4. Experiments and Results

### 4.1 Data Set

We collected up to 6-levels of outlinks from the homepages of 145 animal rights activist groups (e.g., Animal Liberation Front (ALF), People for the Ethical Treatment of Animals (PETA), etc.). The testbed contained 524,483 web pages with a size of about 25 GB.

We aimed to collect content containing positive sentiments towards animal rights and animal protection initiatives (C1). Such content sheds light on an important and active constituency that exerts considerable influence on the political and corporate landscapes. The testbed also contained content with neutral or opposing sentiments (C2), such as objective information and news about these groups, as well as criticism targeted towards animal rights activists by individuals and groups holding opposing views. The variety of content in the test bed makes it suitable for our experiments.

To train our text classifier module and graph comparison module, we built a training data set that consisted of 800 targeted web pages (C1) and 800 irrelevant ones (C2, C3, & C4). These pages were manually selected by two domain experts, based on topic relevance and sentiment relevance, from both our testbed and WWW. Consistent with prior work [4], this data was also used to train an accurate yet computationally expensive gold standard Support Vector Machines classifier that used over 10,000 learned attributes [8, 9]. The classifier, which attained 89.4% accuracy on 2,000 randomly selected testing pages, was applied on the entire testbed. Only 81,370 pages (15.5%) were classified as relevant in both topic and sentiment (C1). We also used public inlink service to collect up to 3-level inlink pages for the 1,600 training pages. The class information of these inlink pages are used to construct the web graphs of our training data, which help to train our graph module.

### 4.2 Experiment 1: Evaluate the usefulness of sentiment information

The first experiment focuses on utilizing outputs of our text classifier module. Breadth First Search (BFS) is used as a benchmark like many previous crawler studies. The following additional variations of BFS are implemented to examine the usefulness of sentiment information in spidering tasks with both topic and sentiment requirements:

BFS-13: BFS that explores sentiment relevant content first (C1 & C3)
BFS-12: BFS that explores topic relevant content first (C1 & C2)
BFS-1: BFS that explore targeted content first (C1)
BFS-1-2: BFS that explore targeted C1 first, C2 second, and C3 & C4 last

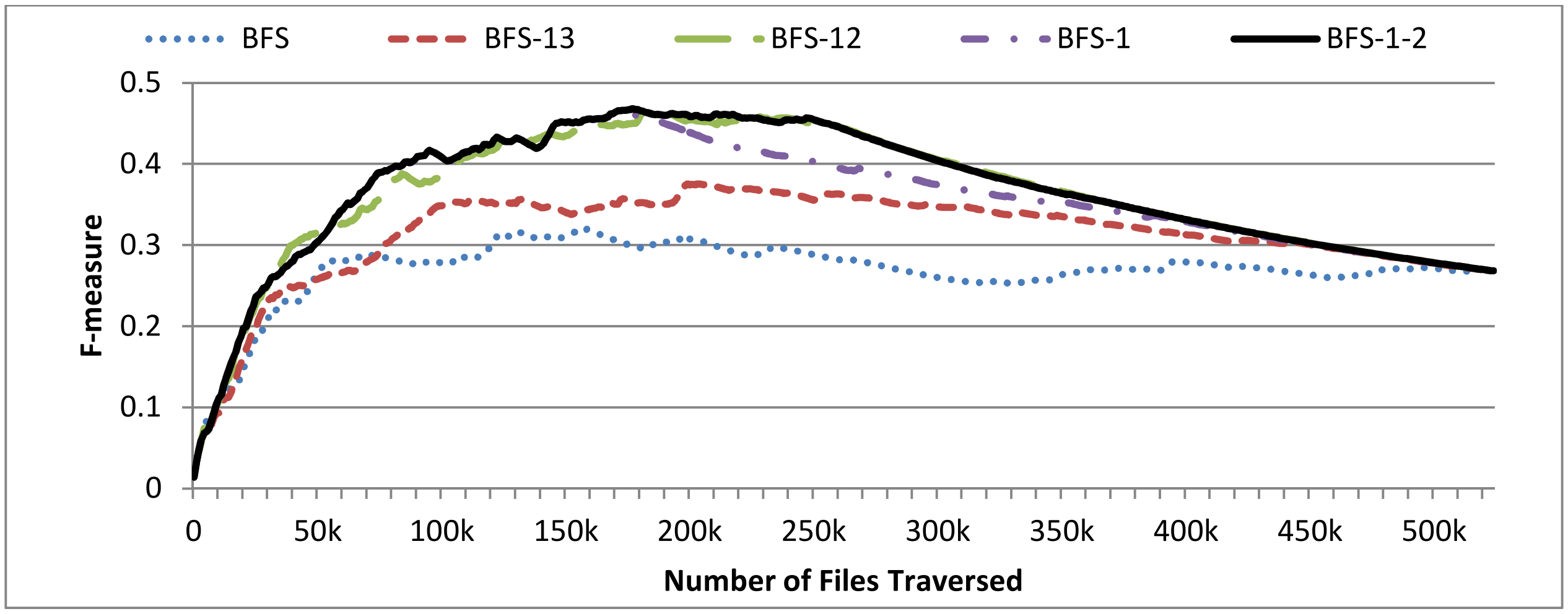

Fig 1. F-Measure Results for Experiment 1

The F-measure results are shown in Figure 1. Providing only sentiment information to BFS (BFS-13) resulted in about 5% increase on F-Measure. Combining sentiment with topic information boosted F-measure by 12-15% in most cases. BFS-1-2 performed best among all these methods. Compared with traditional methods that rely on topic information only (BFS-12), BFS-1-2 increased F-measure by over 3%The results support the notion that sentiment information can improve the performance of focused crawlers seeking to collect opinionated content on a particular topic.

### 4.3 Experiment 2: Evaluate the usefulness of web graph

In the second experiment, we evaluate the usefulness of web graph in learning strategies when relevant contents are near irrelevant pages. Our graph comparison module is applied to determine the weights of off-topic pages such as C3 and C4 based on their web graphs. For C1 and C2, BFS-1-2 described in experiment 1 is still used. After BFS-1-2 explored links of all the collected C1 and C2 pages, there are only 5,400 targeted C1 pages left among the rest 260,000 web pages. These 5,400 C1 pages are linked by off-topic pages and usually ignored by traditional approaches. The percentage of such pages could be larger if our testbed is collected based on a more diverse set of seed URLs. We compare our graph module with BFS-1-2 by counting the difference in numbers of collected C1 pages while exploring the rest 260,000 pages. The results are shown in Figure 2.

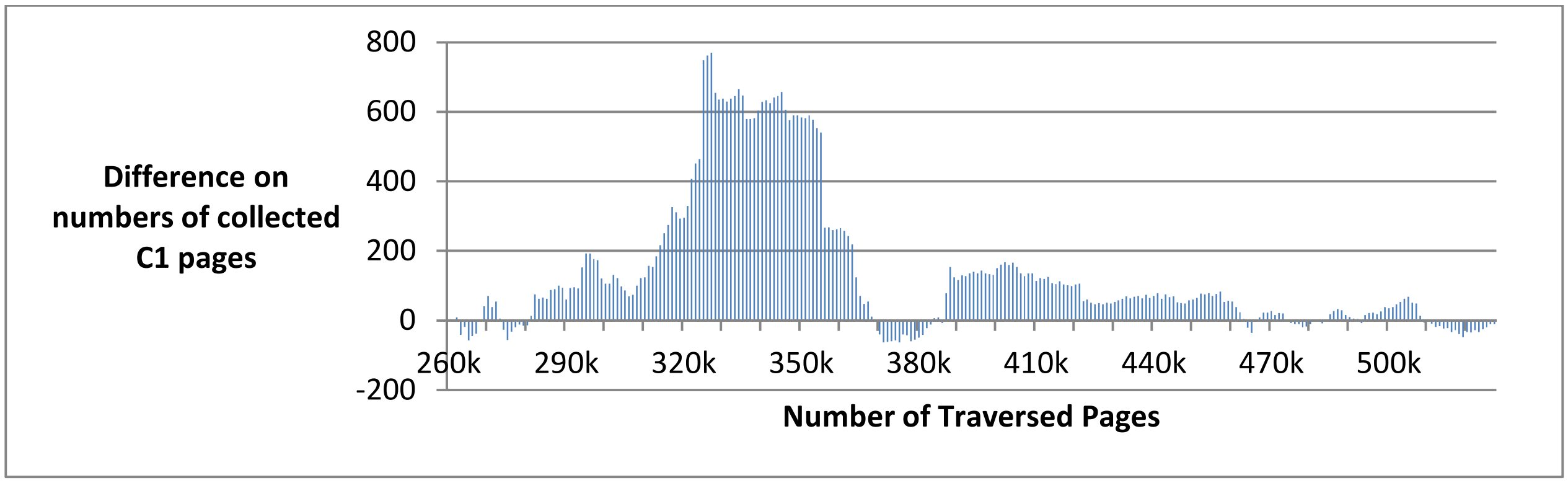

Fig 2. Results for Experiment 2

The vertical axis shows how many more C1 pages our module has collected than the BFS-1-2. It is clear that our graph comparison module outperforms the BFS-based method. After collecting 65,000 pages of the rest 260,000 pages, our module has collected almost 14.8% more C1 pages. There are only a few traversal points when the BFS-based method performs equivalent, or marginally better than the graph comparison module. The experimental result demonstrates that sentiment labelled web graphs can facilitate learning strategies for collecting relevant pages by tunnelling through irrelevant ones.

## 5. Conclusions and Future Directions

In this study, we propose a graph-based sentiment crawler algorithm. Two major contributions of our study are as follows. First, we demonstrate that sentiment is useful for crawling tasks that considers not only topics but also their related sentiments. Second, we provide a new web graph-based method to rank links of off-topic pages. This method helps crawlers learn strategies when relevant contents are near irrelevant pages. In the future, we will focus on further improving the performance of our system by deploying advanced sentiment analysis methods, developing new search algorithms, and tuning various parameters of the graph comparison module.